\documentclass[conference]{IEEEtran}
\usepackage{amsfonts,latexsym,amsmath,amstext,amssymb,verbatim,epsfig,psfrag}
\usepackage{colordvi,pstcol}
\usepackage{graphicx}
\usepackage{psfrag,graphicx}

\newcommand\II{\mathcal{I}}

\def\R{\mbox{I\hspace{-.15em}R}}

\def\R{\mbox{I\hspace{-.15em}R}}

\def\Pb{\bold{P}}

\def\Ib{\bold{I}}
\def\Jb{\bold{J}}

\newtheorem{theorem}{Theorem}

\begin{document}
\title{Fast ranking algorithm for very large data}
\author{\IEEEauthorblockN{Dohy Hong}
\IEEEauthorblockA{Alcatel-Lucent Bell Labs\\
Route de Villejust\\
91620 Nozay, France\\
\{dohy.hong\}@alcatel-lucent.com}
}

\date{\today}
\maketitle

\begin{abstract}
In this paper, we propose a new ranking method inspired from previous results
on the diffusion approach to solve linear equation.
We describe new mathematical equations corresponding to this method and show
through experimental results the potential computational gain.
This ranking method is also compared to the well known PageRank model.
\end{abstract}




\begin{psfrags}
\section{Introduction}\label{sec:intro}
Inspired from the previous research results on the diffusion approach \cite{d-algo}, \cite{cv}
to solve fixed point problem in linear algebra, we propose here a new data ranking definition
and algorithm. This result can be seen as a mix of PageRank solution \cite{page}, diffusion approach
\cite{d-algo} and path diversity idea \cite{diversity}.

In Section \ref{sec:def}, we define the notations and the theoretical framework. 
Section \ref{sec:eval} show the first experimental results, including the comparison
to PageRank ranking.
\section{Algorithm description}\label{sec:def}
\subsection{Notations}

We will use the following notations:
\begin{itemize}
\item $\Pb \in \R^{N\times N}$ a real matrix;
\item $\Ib\in \R^{N\times N}$ the identity matrix;
\item $\Jb_i$ the matrix with all entries equal to zero except for
  the $i$-th diagonal term: $(\Jb_i)_{ii} = 1$;
\item $\Omega = \{1,..,N\}$;
\item $\II = \{i_1,i_2,..,i_n,...\}$ the sequence of nodes for the update
  (diffusion): $i_k \in \Omega$;
\item $E = (1,..,1)^T \in \R^N$;
\item $L_1$-norm: if $X\in \R^N$, $|X| = \sum_{i=1}^N |x_i|$;
\end{itemize}

We assume that $\Pb$ is the matrix associated to the directed graph on $\Omega$,
for instance, the PageRank matrix (i.e. the transition matrix multiplied by the
damping factor \cite{page}).

\subsection{Fast ranking algorithm}
The proposed ranking algorithm is based on the iteration of the double equations
on history $H_n$ and fluid $F_n$ vectors:

\begin{eqnarray}
H_0 &=& 0\nonumber\\
H_n &=& H_{n-1} + \Jb_{i_n} ((int)F_{n-1})\label{eq:defHex}
\end{eqnarray}
and
\begin{eqnarray}
F_0 &=& \alpha . (1,..,1)^T\nonumber\\
F_n &=& F_{n-1} + (\Pb - \Ib) \Jb_{i_n} ((int)F_{n-1}).\label{eq:defFex}
\end{eqnarray}

The above equations can be easily interpreted as: we apply exactly the algorithm of
D-iteration \cite{d-algo}, but we only diffuse the integer part of fluids.
The Jacobi iterations of the above equations are defined by:
\begin{eqnarray*}
H_n &=& H_{n-1} + (int)F_{n-1}\\
F_n &=& F_{n-1} + (\Pb - \Ib) ((int)F_{n-1}).
\end{eqnarray*}

If $\Pb$ is a non negative matrix with spectral radius less than unity, 
it is obvious to see that $F_n$ and $H_n$ converge ($H_n$ is non-decreasing bounded by PageRank vector)
in a finite number of steps (if not, $H_n$ would be unbounded) and the proposed ranking (FR)
method is based on $H_{\infty}+F_{\infty}$. $F_{\infty}$ can be seen as a tie-breaker, but one may also use $H_{\infty}$.
One may also consider a personalized PageRank flavour extension replacing $F_0$
by any other initial vector $V$.
One of this approach's advantage is to be not very dependent on the choice
of the damping factor, both for the ranking and the computation speed (cf. Table \ref{tab:1000}).

To solve the above equations, we will apply the diffusion approach \cite{d-algo}.
This means in particular that this computation method will be naturally suited for the asynchronous
parallel computation, as it was for D-iteration.

Note that if one would diffuse all fluid retained in $F_{\infty}$, $H_{\infty}$ would be
exactly the PageRank vector. The motivations of using $(int)F_{n-1}$ instead of
$F_{n-1}$ are:
\begin{itemize}
\item ranking quality improvement: indeed, as it has been shown in \cite{diversity}, we
  think that the original PageRank vector may be too much influenced by what we could call
  self-estimation. In presence of loops (self-loop or loops of longer length), a
  part of scores that are inherited will be returned to the {\em sender}, which is not
  necessarily the desired property;
\item computation/convergence acceleration cf. Figure \ref{fig:compa};
\item computation/convergence acceleration for the ranking updates when the graph
 (or matrix $\Pb$) evolves in time.
\end{itemize}

If $\alpha$ goes
to infinity, the proposed ranking vector converges to PageRank vector.
Therefore, PageRank can be seen as a particular case $\alpha\to\infty$ of our model
where $\alpha$ tunes the desired influence of loops on the ranking score.
Finally, we have also the following interesting bound on the error:

\begin{theorem}
$\frac{(1-d)}{\alpha N} H_{\infty}$ is an approximation of the PageRank vector with
$L_1$-norm error bounded by $1/(\alpha-1)$.
\end{theorem}
\proof
The proof is based on the monotone property of the diffusion.
If we denote by $H(\alpha E,\beta)$ the limit of FI with initial condition $\alpha E$
and diffusion of the $\beta$-integer part of $F_n$: $(int)(F_n/\beta)\times\beta$, then
we have $H(\alpha E,1) = \alpha H(E,1/\alpha) = \alpha^2 H(E/\alpha E,1/\alpha^2)$ etc.
Let's call $X$ the PageRank vector. 
Then, $X$ can be obtained from the diffusion
of $F(\alpha E,1)$ plus $H(\alpha E,1)$. Note that the diffusion of  $F(\alpha E,1)$
can be denoted by $H(F(\alpha E,1),0)$ (implying $X = H((1-d)/N E,0) = (1-d)/N H(E,0)$).
Now using $H(F(\alpha E,1),0) \le H(E,0)$, we have:
$$
H(\alpha E,1) \le \frac{\alpha N}{(1-d)}X =  H(\alpha E,1) + H(F(\alpha E,1),0)
$$
and
$$
H(\alpha E,1) \le \frac{\alpha N}{(1-d)} X \le H(\alpha E,1) + H(E,0).
$$
Then applying FI in $1/\alpha$-integer part and the same inequality recursively,
we obtain:
\begin{align*}
H(\alpha E,1) \le \frac{\alpha N}{(1-d)} X \le& H(\alpha E,1) + H(E,\alpha^{-1})\\
& + H(\alpha^{-1} E,\alpha^{-2}) + ...
\end{align*}

Therefore
\begin{align*}
H(\alpha E,1) \le \frac{\alpha N}{(1-d)} X \le& H(\alpha E,1)\left(1 + \alpha^{-1} + \alpha^{-2} + ...\right)\\
& = \frac{\alpha}{\alpha-1}H(\alpha E,1)
\end{align*}
And
\begin{align*}
0 \le \frac{\alpha N}{(1-d)} X - H(\alpha E,1) \le& \frac{1}{\alpha-1}H(\alpha E,1)
\end{align*}
which can be rewritten as:
\begin{align*}
0 \le X - \frac{(1-d)}{\alpha N}H(\alpha E,1) \le& \frac{1}{\alpha-1} \frac{(1-d)}{\alpha N} H(\alpha E,1).
\end{align*}
Since $|H(\alpha E,1)|\le \alpha N /(1-d)$, we have:
\begin{align*}
\left|X - \frac{(1-d)}{\alpha N}H(\alpha E,1)\right| \le& \frac{1}{\alpha-1}.
\end{align*}

This means that choosing $\alpha=1000$ would gives an error very close to $0.1$\% for norm $L_1$
but also for each coordinate (the exact bound is $1/999=0.001001001...$).

Note that we also have:
\begin{align*}
&\left| X - \frac{(1-d)}{\alpha N}(H(\alpha E,1)+F(\alpha E,1))\right|\\
& \le\frac{1}{\alpha-1} - \frac{1-d}{N}|F(E,\alpha^{-1})|.
\end{align*}

\section{Experimental evaluation}\label{sec:eval}

For the experimental evaluation purpose,
we took the web graph imported from the dataset \verb+uk-2007-05+ \verb+@1000000+
(available on \cite{webgraphit}) which has 41,247,159 links on $10^6$ nodes.

Below we vary $N$ from $10^3$ to $10^6$ extracting from the dataset the
information on the first $N$ nodes.
Few graph properties are summarized in Table \ref{tab:1}:
\begin{itemize}
\item L: number of non-null entries (links) of $P$;
\item D: number of dangling nodes (0 out-degree nodes);
\item E: number of 0 in-degree nodes: the 0 in-degree nodes are defined recursively:
  a node $i$, having incoming links from nodes that are all 0 in-degree nodes, is
  also a 0 in-degree node; from the diffusion point of view, those nodes are those
  who converged exactly in finite steps;
\item O: number of loop nodes ($p_{ii} \neq 0$);
\item $\max_{in} = \max_i \#in_i$ (maximum in-degree, the in-degree of $i$ is the number of
  non-null entries of the $i$-th line vector of $P$);
\item $\max_{out} = \max_i \#out_i$ (maximum out-degree, the out-degree of $i$ is the number of
  non-null entries of the $i$-th column vector of $P$).
\end{itemize}

\begin{table}
\begin{center}
\begin{tabular}{|l|cccccc|}
\hline
N        & L/N  & D/N   & E/N   & O/N & $\max_{in}$ & $\max_{out}$\\
\hline
$10^3$   & 12.9 & 0.041 & 0.032 & 0.236 & 716   & 130\\
$10^4$   & 12.5 & 0.008 & 0.145 & 0.114 & 7982  & 751\\
$10^5$   & 31.4 & 0.027 & 0.016 & 0.175 & 34764 & 3782\\
$10^6$   & 41.2 & 0.046 & 0     & 0.204 & 403441& 4655\\
\hline
\end{tabular}\caption{Extracted graph: $N=10^3$ to $10^6$.}\label{tab:1}
\end{center}
\end{table}

Table \ref{tab:1000} shows the comparative evaluation of the computation cost in number of
iterations (one iteration is here defined as a use of $L$ coordinates of $\Pb$ in the computation)
with a target precision of $1/N$ (for $L_1$ norm).
We compared the Jacobi iteration, D-iteration (DI, cf. \cite{d-algo}) and the fast ranking algorithm
(FI) we propose in this paper. The convergence becomes very slow when the damping factor $d$ is being
close to 1 to compute the PageRank vector, whereas our ranking vector can be obtained very
efficiently whatever the choice of $d$.
\begin{table}
\begin{center}
\begin{tabular}{|l|ccccc|}
\hline
d        & Jacobi  & DI   & FI ($\alpha=1$) & FI ($\alpha=2$) & FI ($\alpha=10$)\\
\hline
0.85     & 26      & 12   & 1.72 & 3.12 & 6.97\\
0.9      & 36      & 17   & 1.99 & 3.94 & 9.55\\
0.99     & 330     & 101  & 3.60 & 19.6 & 53.1\\
0.999    & 5076    & 548  & 15.3 & 92.7 & 258\\
\hline
\end{tabular}\caption{Computation cost: $N=10^3$. Impact of damping factor. Computation cost: number of use of coordinates of $\Pb$ divided by $L$.}\label{tab:1000}
\end{center}
\end{table}

Figure \ref{fig:compa} shows the convergence speed (in number of iterations) of Jacobi, D-iteration (DI) and
the proposed (FI) methods: our approach reaches the limit in all cases in less than 2.2 iterations.
The convergence efficiency is simply not comparable.
\begin{figure}[htbp]
\centering
\includegraphics[angle=-90,width=8cm]{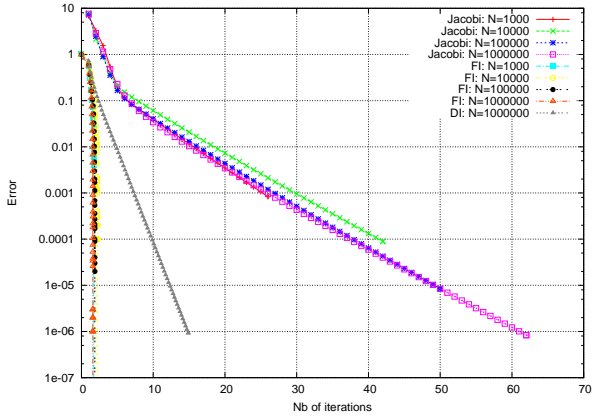}
\caption{Convergence speed comparison.}
\label{fig:compa}
\end{figure}

Figures \ref{fig:top3} and \ref{fig:top6} compare the ranking results obtained with FR (using $H+F$), 
LOC (local computation: rank equal to the number of incoming links) to PageRank vector on the top $x$\%: 
on y-axis, it counts the number of common nodes in the top $x$\% between two ranking methods, then it is divided by
the number of compared elements (nodes). 
The fifth curve shows the common elements proportion we observe between two PageRank vectors using a damping factor of $0.9$ and $0.8$.
Our FR ranking vector can be seen as an approximation of PageRank vector, since it converges to PageRank vector
for large $\alpha$: however, by its definition,
it tends to eliminate self-ranking aspects due to the presence of loops (a part of scores that I give to children nodes
is coming back to me). Therefore, the parameter $\alpha$ is meant to tune the influence of loops in the ranking score
and PageRank can be seen as a particular case $\alpha\to\infty$.
Globally, we see that our ranking vector preserves very well the top ranked web sites
(for $N=10^6$, we see that $\alpha=2$ is close enough already to PageRank vector, always above 92\%), because they
are likely to be pointed by many different and relevant other web sites: FR ranking vector
includes by its definition features and ideas of the path diversity mechanism proposed in \cite{diversity},
when $\alpha$ is closer to 1, but with a computation cost that is greatly reduced (whereas the ideas in \cite{diversity}
requires more computation cost than PageRank vector computation).

Even though it is hard to justify theoretically, the author believe that a choice of $\alpha$
between 1 and 2 are the most appropriate in terms of the optimal compromise between computation
cost and ranking relevancy.


\begin{figure}[htbp]
\centering
\includegraphics[angle=-90,width=8cm]{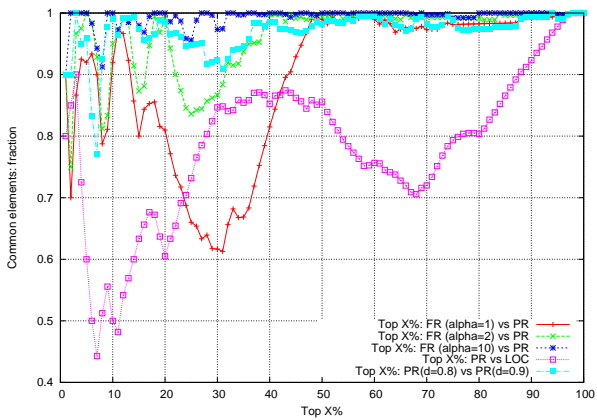}
\caption{$N=10^3$: proportion of common elements between FR (proposed) and PR (PageRank) in top $x$\% ranked sites.}
\label{fig:top3}
\end{figure}

\begin{figure}[htbp]
\centering
\includegraphics[angle=-90,width=8cm]{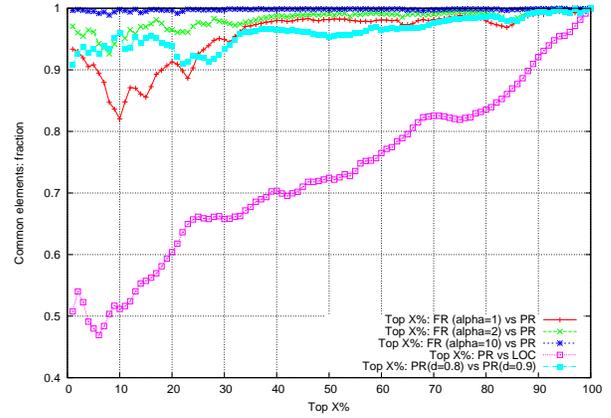}
\caption{$N=10^6$: proportion of common elements between FR (proposed) and PR (PageRank) in top $x$\% ranked sites.}
\label{fig:top6}
\end{figure}

\section{Conclusion}\label{sec:conclusion}
In this paper, we proposed a new data ranking method and compared its efficiency
to the computation of PageRank vector. Applying the diffusion method on this
new ranking vector, we showed that a very efficient computation can be obtained
while targeting a relevant ranking score as PageRank.

\end{psfrags}
\bibliographystyle{abbrv}
\bibliography{sigproc}

\begin{thebibliography}{1}

\bibitem{webgraphit}
http://law.dsi.unimi.it/datasets.php.

\bibitem{d-algo}
D.~Hong.
\newblock D-iteration method or how to improve gauss-seidel method.
\newblock {\em arXiv, http://arxiv.org/abs/1202.1163}, February 2012.

\bibitem{diversity}
D.~Hong.
\newblock Statistical reliability and path diversity based pagerank algorithm
  improvements.
\newblock {\em arXiv, http://arxiv.org/abs/1202.2393}, Feb 2012.

\bibitem{cv}
D.~Hong, F.~Mathieu, and G.~Burnside.
\newblock Convergence of the d-iteration algorithm: convergence rate and
  asynchronous distributed scheme.
\newblock {\em arXiv, http://arxiv.org/abs/1301.3007}, January 2013.

\bibitem{page}
L.~Page, S.~Brin, R.~Motwani, and T.~Winograd.
\newblock The pagerank citation ranking: Bringing order to the web.
\newblock {\em Technical Report Stanford University}, 1998.

\end{thebibliography}

\end{document}